\documentclass[aps,twocolumn,showpacs,nofootinbib]{revtex4-2}
\usepackage{graphicx}
\usepackage{graphics}
\usepackage{textcomp}
\usepackage{psfrag}
\usepackage{epstopdf}
\usepackage{amssymb}
\usepackage{amsmath}
\usepackage{xcolor}
\usepackage{float}
\usepackage{lineno}
\usepackage{makecell}
\usepackage{setspace}
\usepackage{lipsum}%
\usepackage{inputenc}
\usepackage{tabularx}
\usepackage{soul,collcell,datatool}
\usepackage[colorlinks=true,linkcolor=blue,citecolor=blue,urlcolor=blue]{hyperref}
\usepackage[normalem]{ulem}

\begin{document}
\setstretch{1.0}
\renewcommand\linenumberfont{\normalfont\scriptsize\sffamily\color{red}}

\title{Unraveling the electronic, vibrational, thermodynamic, optical and piezoelectric properties of LiNbO$_3$, LiTaO$_3$ and Li$_2$NbTaO$_6$ from first-principles calculations} 

\author{Debidutta Pradhan, Rojalin Swain, Souvagya Kumar Biswal and Jagadish Kumar$^{}$}
\email{Corresponding author	jagadish.physics@utkaluniversity.ac.in}
\affiliation{Center of Excellence in High Energy and Condensed Matter Physics, Department of Physics, Utkal University, Bhubaneswar 751004, India 
}
\begin{abstract}
We have investigated the electronic, vibrational, optical, thermal and piezoelectric properties of LiNbO$_3$, LiTaO$_3$ and Li$_2$NbTaO$_6$ using the first-principles calculation based on the density functional theory. It also shows structural phase transition below $T_c$ due to ionic displacement that may alter the properties of material. We have checked the structural stability by calculating the tolerance factor and formation energy before proceeding to the further calculations. The ground state electronic band structures and corresponding density of states show its semiconducting nature with a band gap range of 3.5-3.7 eV. Optical properties such as dielectric function, absorption coefficient, optical conductivity, refractive index, absorbance and reflectance are calculated using time-dependent density functional theory. Furthermore, the piezoelectric properties and Born effective charges were analyzed to find the correlation between them.  In these materials, the distortion induced by the small ionic radius of Li$^{+}$ coupled with strong covalent interaction between transition metal and oxygen leads to high spontaneous polarization which can  enhance both piezoelectric and optical properties.
\end{abstract}

\maketitle
\section{Introduction} 
Piezoelectricity is a fundamental property of certain crystalline materials which has diverse applications in sensors, actuators, and energy harvesting devices. The discovery and utilization of piezoelectric materials have led to advancements in high-precision instrumentation, sonar technology, and even wearable devices for self-powered electronics \cite{Martin72,Safari08,Uchino17}. Historically, the most widely used piezoelectric materials have been lead-based perovskites, particularly lead zirconate titanate (PZT), owing to their high piezoelectric coefficients, Curie temperature and stability \cite{Gubinyi08,Zou23,Huang23}. However, due to environmental concerns and stringent regulations associated with lead toxicity, there has been a persistent drive towards the development of lead-free alternatives \cite{EUdir02,EUdir11,Rodel15,Wu15}. The search for lead-free perovskites has gained significant momentum in the past two decades, with exploration of various classes of materials that could match or even surpass the performance of lead-based counterparts. Several types of lead-free perovskites have been discovered and synthesized over time, including bismuth-based perovskites (BiFeO3), alkali niobates (for example potassium sodium niobate) and double perovskites with mixed cations and anions to enhance structural and electronic properties. Notable progress has been made in optimizing these materials for practical applications, with extensive research focusing on enhancing their piezoelectric response, stability, and manufacturability \cite{Acosta17}. Alkaline niobate was initially considered a promising alternative to lead-based perovskites, however, its efficiency is relatively low, and it exhibited limited performance at lower operating temperatures \cite{Saito04}. Significant improvements were later achieved by incorporating alkaline bismuth and titanium composites, which have enhanced efficiency and better functional properties \cite{Takenaka91,Kanie11,kobayashi04}. Among the promising lead-free piezoelectric materials, lithium niobate (LiNbO3) and lithium tantalate (LiTaO$_{3}$) stand out due to their exceptional piezoelectric and ferroelectric properties. These materials possess a non-centrosymmetric crystal structure that facilitates strong piezoelectric coupling, making them highly effective for use in optical modulators, surface acoustic wave (SAW) devices, and electro-optic applications \cite{Li23,Volk08,Bartasyte17,Kudryashov22}. These characteristics are highly desirable for applications in laser technology, nonlinear optics, and integrated photonic circuits. Their strong electro-optic coefficients, low optical loss and spontaneous polarization render them indispensable in the development of next-generation photonic devices. Because of their unique crystal structures, characterized by distorted octahedral coordination and strong ion displacement, they generate high piezoelectric response. Moreover, the presence of lithium ions in these structures facilitates enhanced domain wall motion, improving the overall electromechanical coupling efficiency \cite{Singh22}.

 In recent years, there has been growing interest in the development of novel double perovskites, such as lithium niobium tantalate (Li$_2$NbTaO$_6$), which exhibit piezoelectric and optical properties with low dielectric losses. This emerging material aims to incorporate the characteristics of both LiNbO3 and LiTaO$_{3}$ while mitigating their respective limitations \cite{Bernhardt24.1,Bartasyte19}. It has potential in tunable photonic and acoustic devices which makes it a highly desirable material for industrial applications like telecommunications, energy harvesting, and biomedical sensors 
 \cite{Roshchupkin20,Suhak21}.
\begin{figure}
\includegraphics[height=5.0cm,width=8.0cm, angle =0]{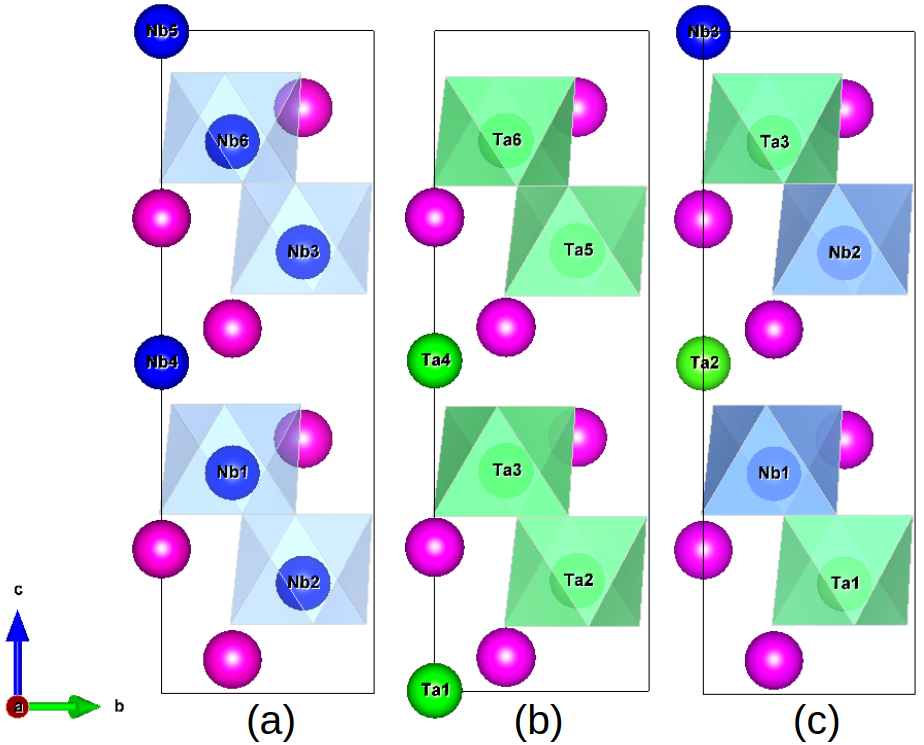}
\caption{Geometrical structures of (a) LNO, (b) LTO and (c) LNTO}
\label{structure}
\end{figure}         
Li$_2$NbTaO$_6$ crystal structure has been designed with space group symmetry $R3c$ and its properties have been systematically investigated using first-principles calculations alongwith LiNbO$_3$ and LiTaO$_{3}$. 
Extensive research is currently underway with the doping of transition metals (e.g., Ti, V, Zn, Ag) and rare-earth metals (e.g., Er) in both bulk and layered structures to enhance functional properties \cite{Theimann94,Fan19,Bridges12,Dai21,Wang22}. 
These advancements align with the growing interest in optoelectronic applications driven by recent breakthroughs in the field of photonics \cite{Kim24}. The integration of these materials into next-generation electronic and photonic systems is expected to drive significant advancements and solidify their importance in modern science and engineering. In this work, we have used the density functional theory to present the electronic structures, phonon dispersion, thermodynamics, optical, piezoelectric properties for possible applications.

\section{Computational Method}
First-principles calculations were performed within the framework of density functional theory(DFT) to obtain the various properties of the compounds. Li$_2$NbTaO$_6$ is designed using VESTA software by taking into account the structures of LiTaO$_{3}$ and Li$_2$NbTaO$_6$  \cite{Matpro,VESTA}. For brevity, we abbreviate LiNbO$_{3}$, LiTaO$_{3}$ and Li$_2$NbTaO$_6$ as LNO, LTO and LNTO respectively. LNTO is designed by substituting Ta alternatively with Nb in LNO structure along the direction $(0,0,1)$. The structures are optimized using high-precision non-magnetic calculations with an increased planewave cutoff energy corresponding to actual k-spacings of 0.495 per angstrom. Electronic band structures and the corresponding density of states, as well as the optical properties, were calculated on a 6x6x6 mesh using a Gaussian smearing of 3000 energy grid points. 
Piezoelectric coefficients for both clamped and relaxed ions along with Born effective charges (BEC) were simulated using the linear-tetrahedron method with Bloechl corrections. All the simulations were performed using the Vienna Ab initio Simulation Package (VASP) with GGA-PBEsol pseudopotential for efficient results \cite{MedeA,vasp1,vasp2,vasp3}.

\section{Results and Discussion}
It is crucial that composites designed from perovskites maintain their structural stability with the same $R3c$ crystal symmetry. We have calculated the tolerance factor ($t = 0.75$) which is $< 1$ and it is same for all the three structures because of the equal ionic radius of Nb and Ta \cite{debi2025}. As per the formulation by Bartel et al., which introduces the oxidation states, the value of tolerance factor is, $\tau=8.06$ showing greater distortion which indicates the materials' stability for off-cubic structure \cite{Bartel19}. 
Smaller size of Li atom  along with strong covalency in Nb-O and Ta-O, leads to octahedral tilting inducing off-centre displacements in oxygen octahedra. This displacement breaks the inversion symmetry inducing better piezoelectric and non-linear optical properties. 
The calculated formation energies per atom are $-2.697, -2.81$ and $-2.738$ eV/mol for LNO, LTO and LNTO respectively confirming the stability of the compound.
\begin{figure}
\includegraphics[height=4.5cm,width=8cm, angle =0]{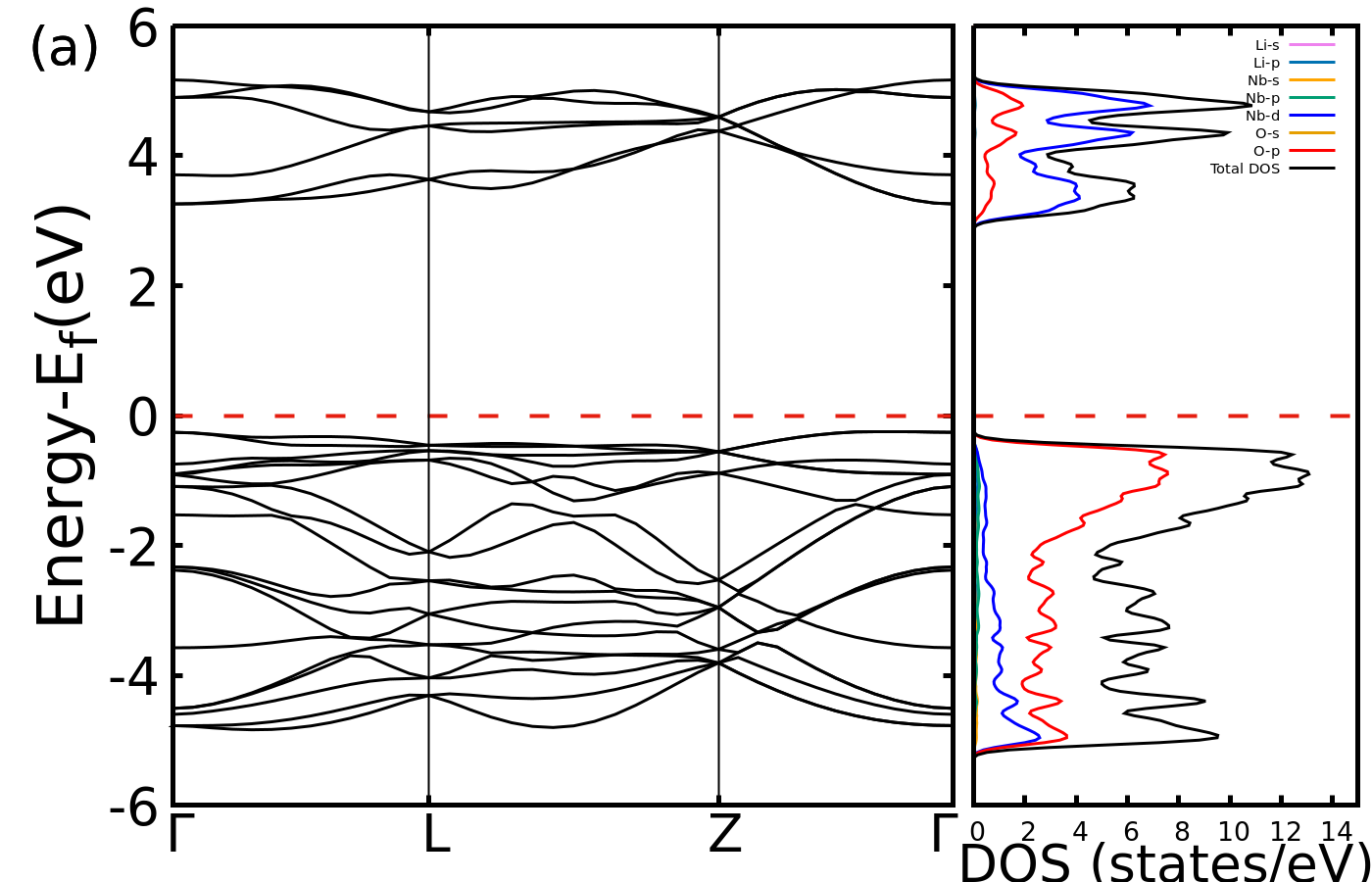}
\includegraphics[height=4.5cm,width=8cm, angle =0]{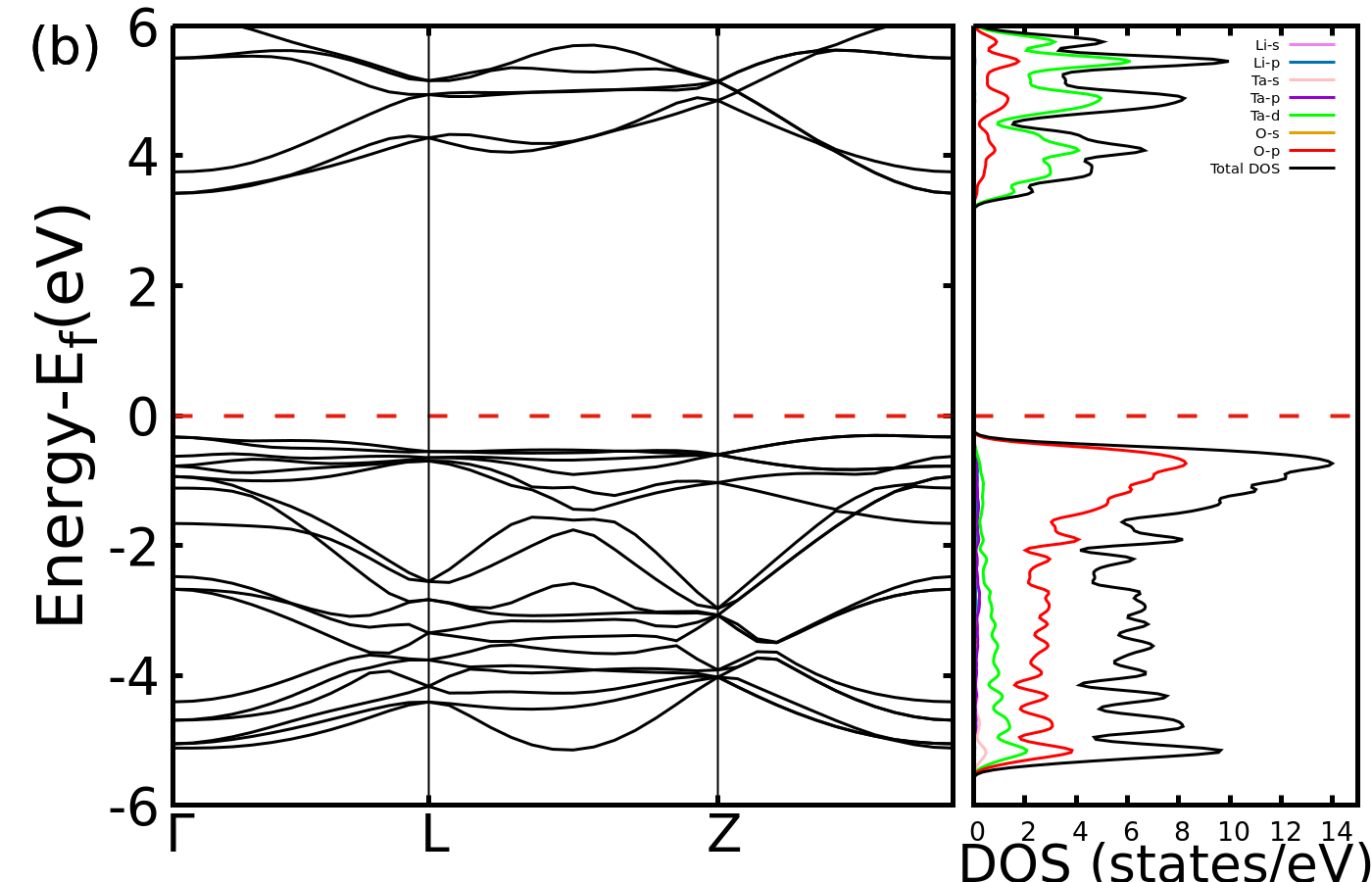}
\includegraphics[height=4.5cm,width=8cm, angle =0]{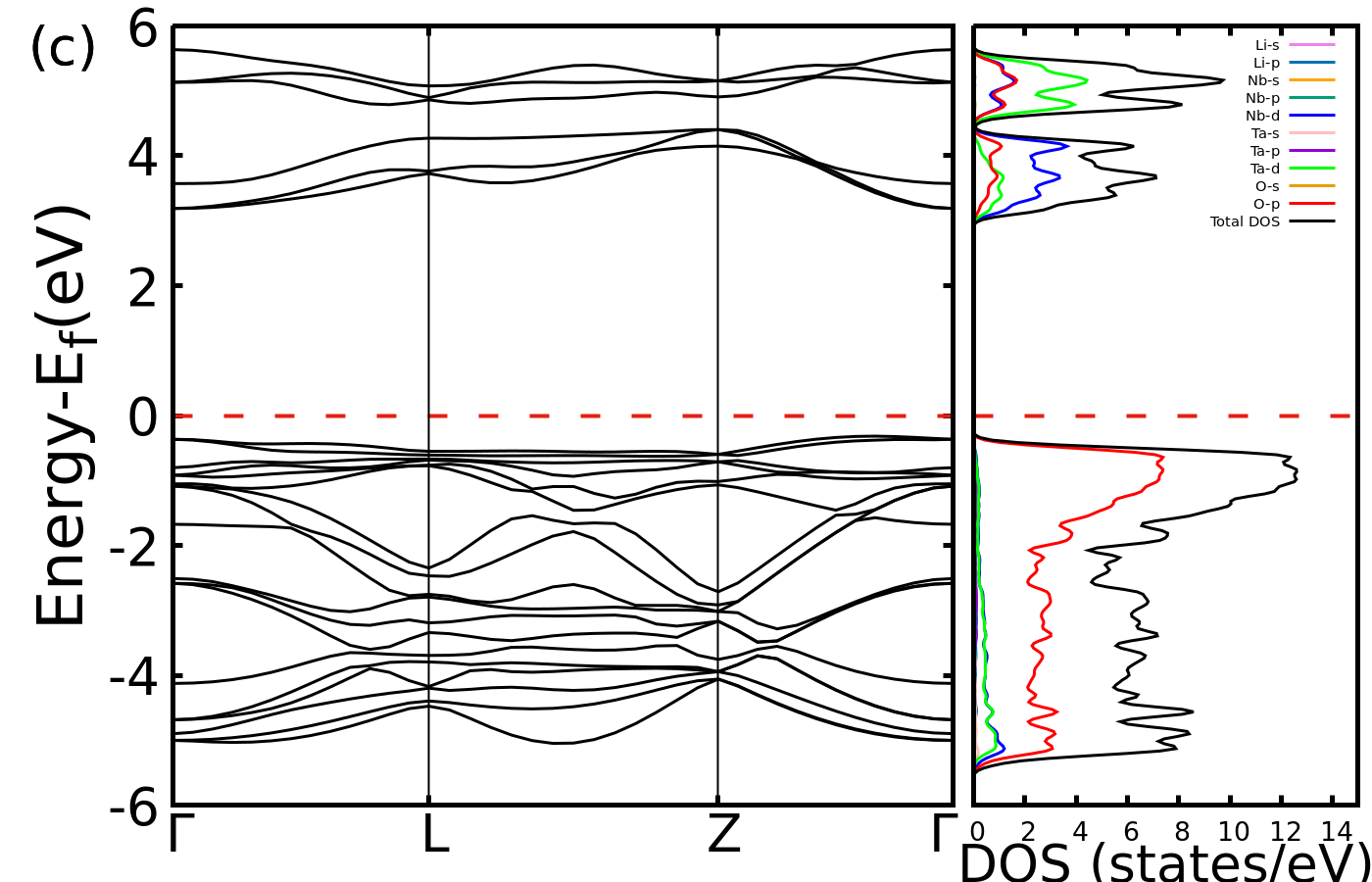}
\caption{Electronic band structures and corresponding density of states for (a) LNO, (b) LTO and (c) LNTO}
\label{band_structure}
\end{figure} 
\subsection{Electronic structure}

The bandgap ($E_g$) is a key characteristic of semiconductors that influences their electrical and optical properties. The band structures and density of states are shown in Fig. \ref{band_structure} having $E_g$, 3.502, 3.728 and 3.550 eV for LNO, LTO and LNTO respectively \cite{Li14,Bernhardt24.1}. The top of the valence band (VB) is dominated by O-$2p$ orbital, while the bottom of the conduction band (CB) with transition metal $d$ orbitals. The energy band gap is influenced by $p-d$ hybridization, which is stronger when the $d$ orbitals are spatially closer in energy to the $p$-orbital. This enhanced hybridization results in a smaller band gap in LNO as compared to LTO. Since LNTO is a combination of Nb$-4d$ and Ta$-5d$ orbitals hybridized with O$-2p$ orbitals, its band gap lies between LNO and LTO. The Nb$-4d$ orbital  contribution is higher as compared to Ta$-5d$ in the conduction band near the Fermi level in LNTO.  The spatial separation of these orbitals inhibits their mixing, resulting in weak interaction between Nb and Ta in the CB. Two separate non-overlapping states in CB could also indicate the allowed optical transitions into different conduction band minima. Valence band is composed of highly hybridized orbitals than the CB and a strong interaction between Nb$-4d$ and Ta$-5d$ orbitals due to the significant contribution of O$-2p$ orbitals and such hybridization is less prominent in the conduction band. The bandgap range is suitable for the UV optoelectronic application, high-power and high-frequency electronics, transparent electronics and displays along with radiation sensors.

\begin{figure}
\includegraphics[height=4.5cm,width=8cm, angle =0]{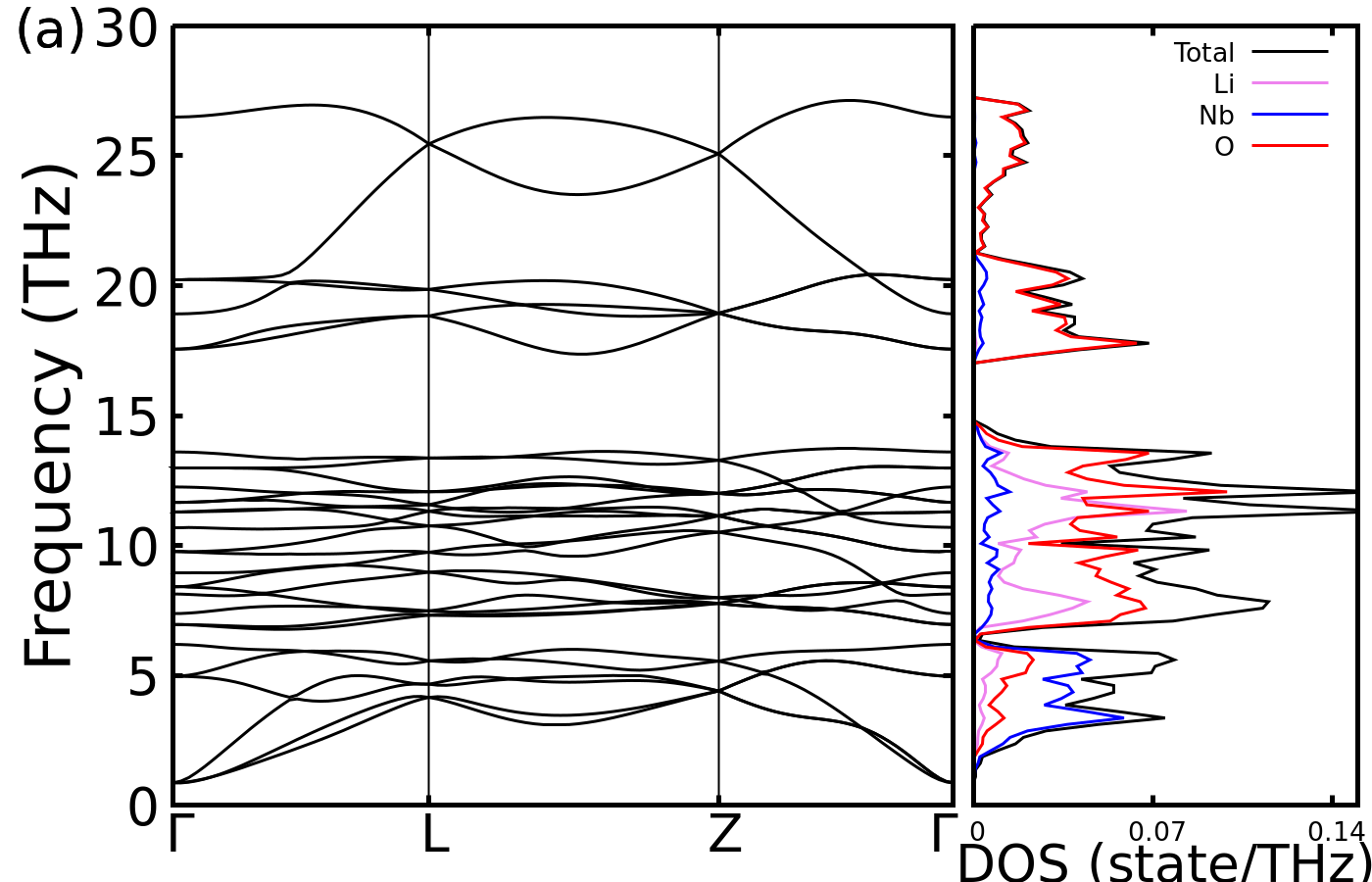}
\includegraphics[height=4.5cm,width=8cm, angle =0]{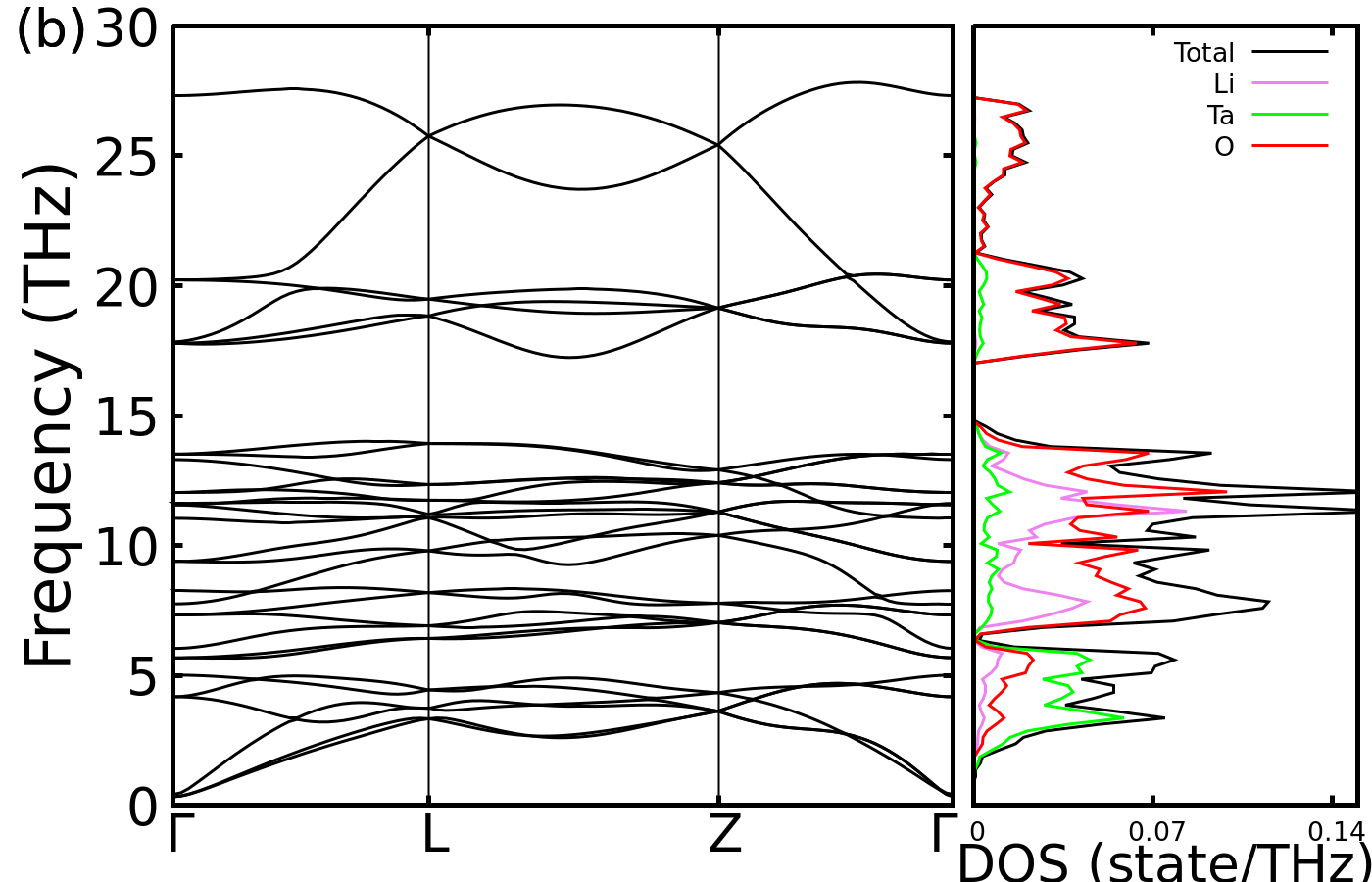}
\includegraphics[height=4.5cm,width=8cm, angle =0]{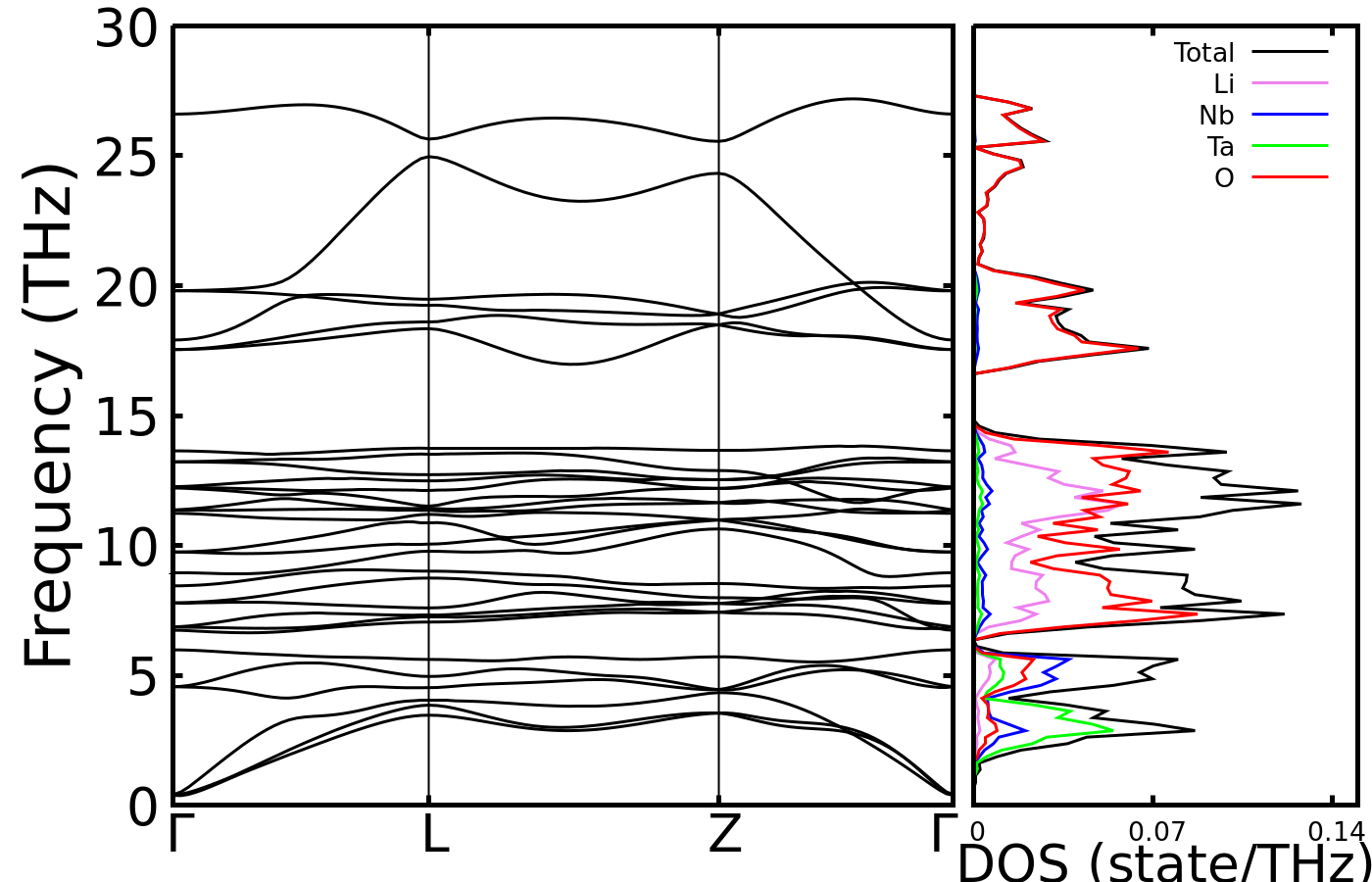}
\caption{Phonon dispersion and density of states of (a) LNO, (b)LTO and (c) LNTO}
\label{phband}
\end{figure}
\subsection{Phonon and thermodynamic properties}
The phonon dispersion curves along the high-symmetry lines in the Brillouin zone, along with the phonon density of states (PDOS) are plotted in Fig. \ref{phband}. Absence of imaginary mode demonstrating the structural stability. Thirty vibrational phonon modes ($27$ optical and $3$ acoustic) are present here because the cell contains $10$ atoms. In Li$_2$NbTaO$_6$, acoustic phonon modes are predominantly influenced by Ta atom because of heavy atomic mass as compared to other element present in the cell, while Nb atom contribute to the lower-frequency optical modes and the high-frequency optical modes are  associated with O and Li atoms. Optical phonon mode separation is due to different masses of niobium, oxygen, as well as Lithium. The contributions of each element to different modes can be estimated from the density of state of each atom.

\begin{figure}
\includegraphics[height=5cm,width=8cm, angle =0]{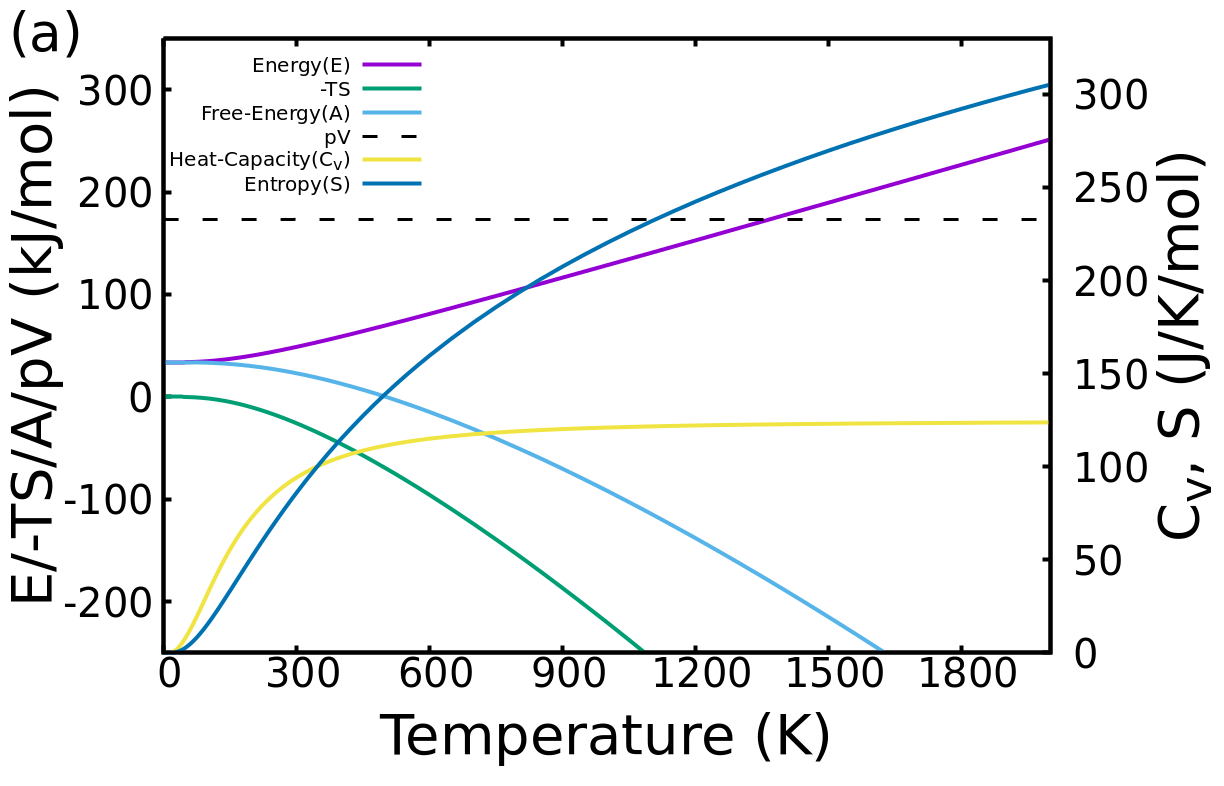}
\includegraphics[height=5cm,width=8cm, angle =0]{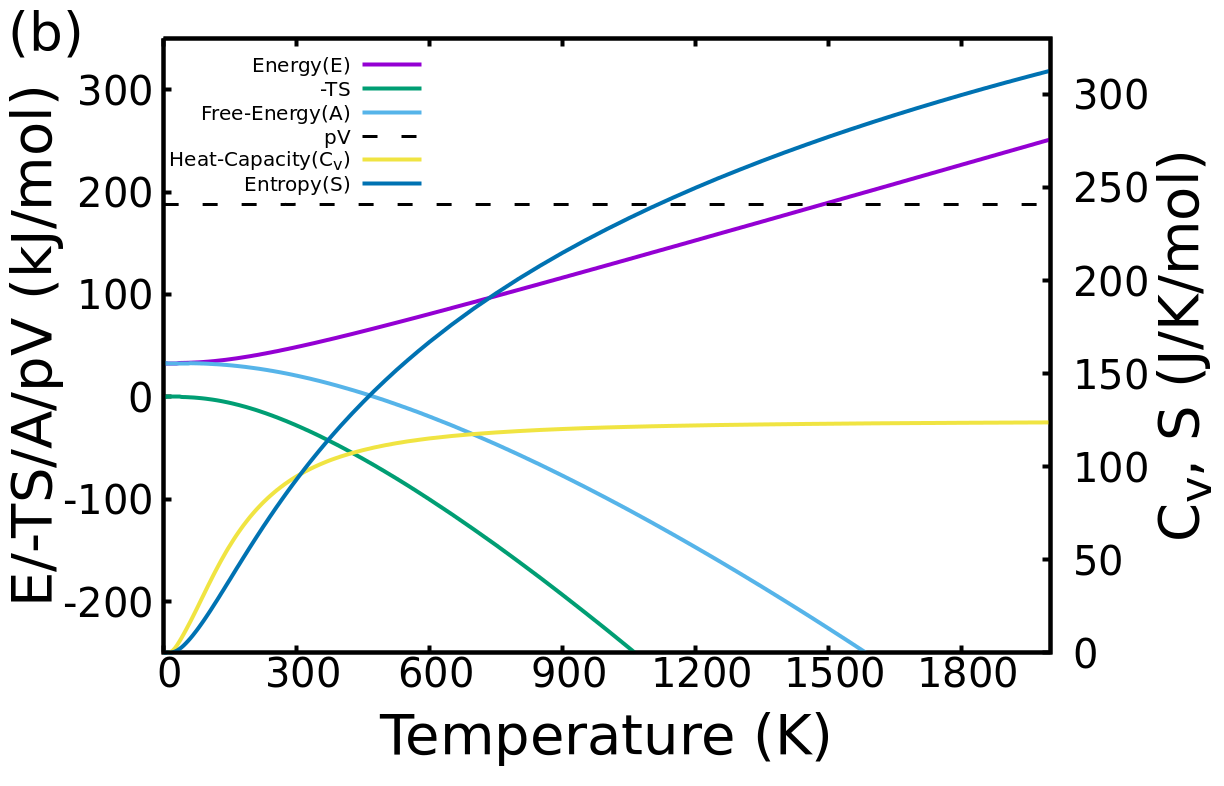}
\includegraphics[height=5cm,width=8cm, angle =0]{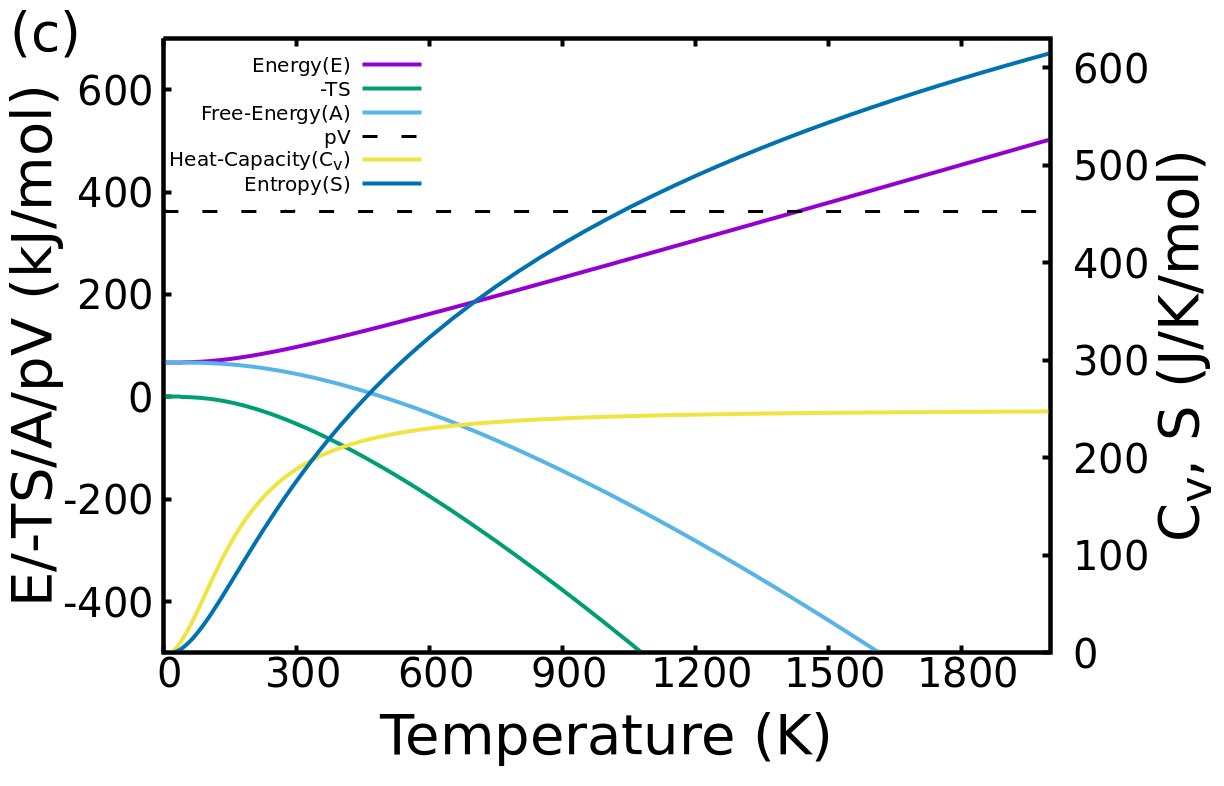}
\caption{Thermodynamic functions of (a) LNO, (b) LTO and (c) LNTO}
\label{band}
\end{figure}

Internal energy, enthalpy, Helmholtz and Gibbs free energies provide a means to assess the stability, phase behaviour and the response to an external perturbation in a system. 
The derivatives of these potentials provide the direct means to calculate other thermodynamic quantities such as heat capacity, chemical potential, entropy, pressure, volume and temperature. 
By calculating the harmonic phonon energy ($E$), the specific heat capacity ($C_v$) can be derived and the partition function that describes the distribution of the particles is used for the calculation of Helmholtz free energy ($A$) and entropy ($S$).
As is evident from the plot, the harmonic phonon energy of the system at absolute zero is 33.77 kJ/mol for LNO, 33.00 kJ/mol for LTO, 67.17 kJ/mol for LNTO which is the zero point energy arising from quantum mechanics. 
This energy consists of residual kinetic and potential energy contributions from microscopic particles. As can be observed, the internal energy starts out with an exponential growth for a small range of temperature before beginning to grow linearly with temperature.
As the internal energy increases linearly, the heat capacity remains constant over the subsequent temperature range. At low temperatures, its exponential growth adheres to Debye’s $T^3$ law, while at high temperatures, $C_V$ asymptotically approaches $3Nk_B$ in accordance with the Dulong-Petit law.
At absolute zero, the system has zero entropy which gradually climbs with increase in temperature. 
The free energy reaches zero at 500 K for LNO, 475 K for LTO, 493 K for LNTO and continues to decrease thereafter, indicating increased system stability, as stability correlates with a reduction in free energy  \cite{Tolborg22}. The change in Helmholtz energy during a thermodynamic process represents the maximum work the system can perform under isothermal conditions. 
At room temperature ($\sim$300 K), 
the free energies per atom of LNO, LTO and LNTO is 4.6 kJ/mol, 4.138 kJ/mol, and 4.46 kJ/mol respectively. This suggests that the structure of LTO is most stable, followed by LNTO and LNO. 

\subsection{Optical properties}
\begin{figure}[!h]
\includegraphics[height=5cm,width=7cm]{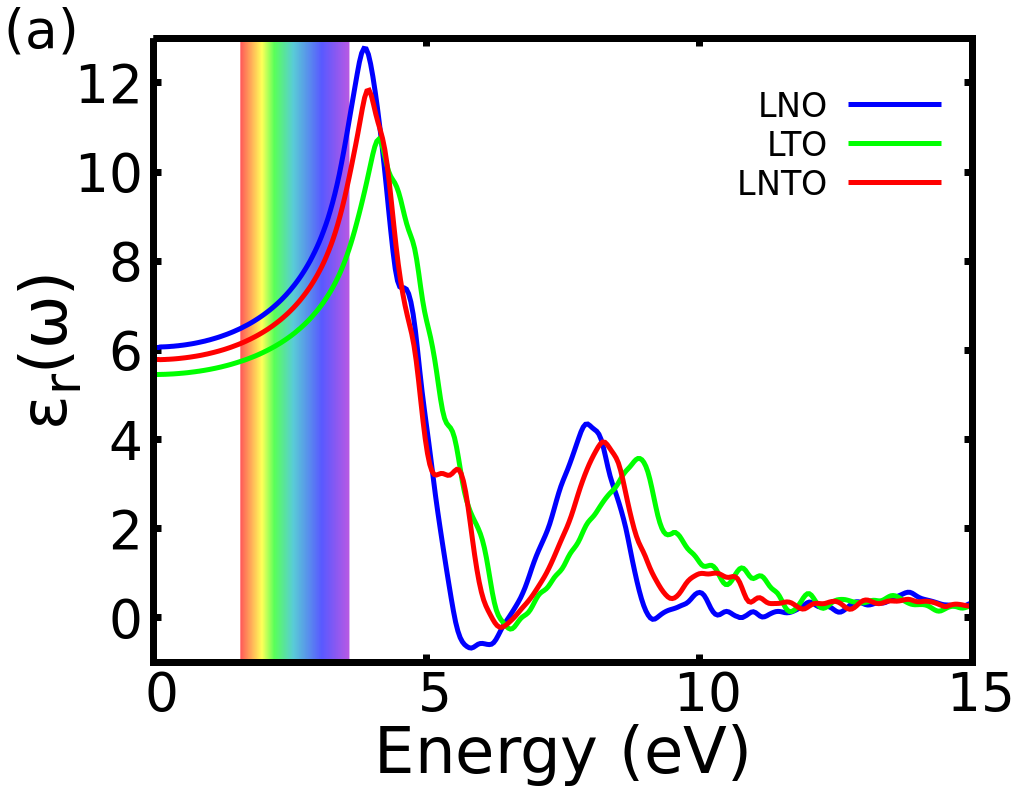}
\includegraphics[height=5cm,width=7cm]{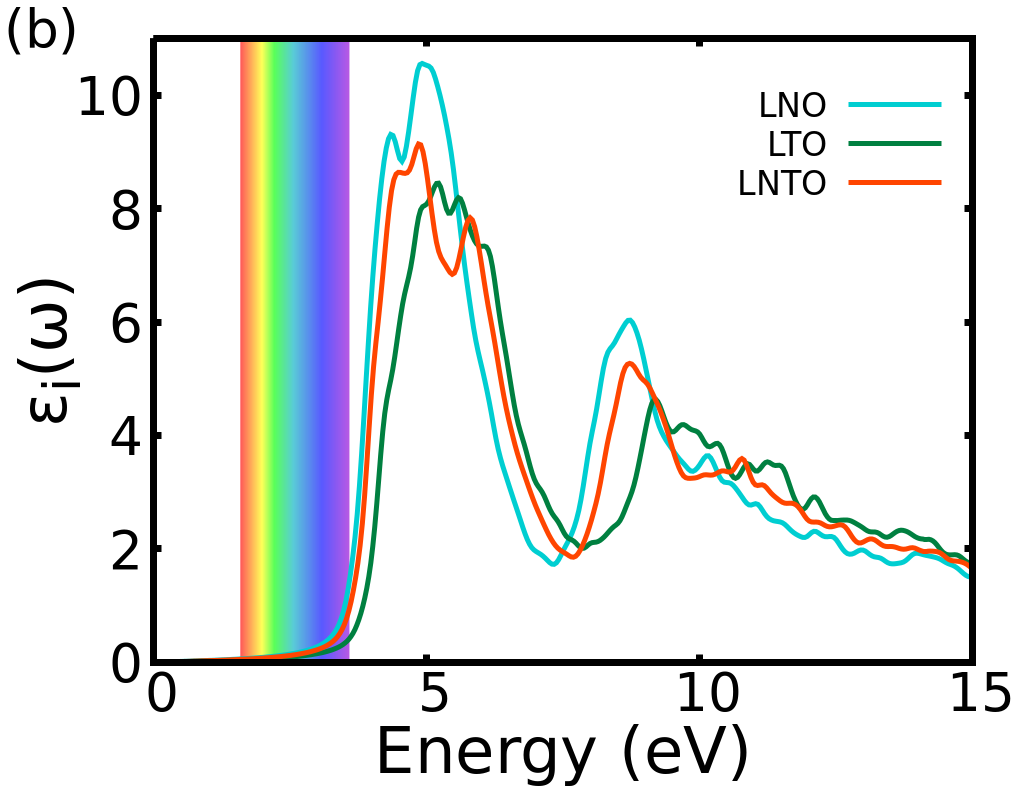}
\caption{Real and Imaginary dielectric properties of (a)Real part and (b)Imaginary part of Dielectric function}
\label{optic_dielectric}
\end{figure}

\indent In the context of optical properties of materials, the dielectric constant and
refractive index are two important parameters that describe how light interacts
with a material. The dielectric constant (also called the relative permittivity) is a measure of how much a material can get polarized in response to an applied electric field, which affects its interaction with electromagnetic waves.
The real and imaginary parts of dielectric function are shown in the Fig. \ref{optic_dielectric}. The imaginary part corresponds to the absorption of the electromagnetic radiation by the material representing the energy loss per unit volume. It is related to the electronic transition between the occupied and the unoccupied states of the material.
\begin{figure}[!h]
\includegraphics[height=5cm,width=7cm]{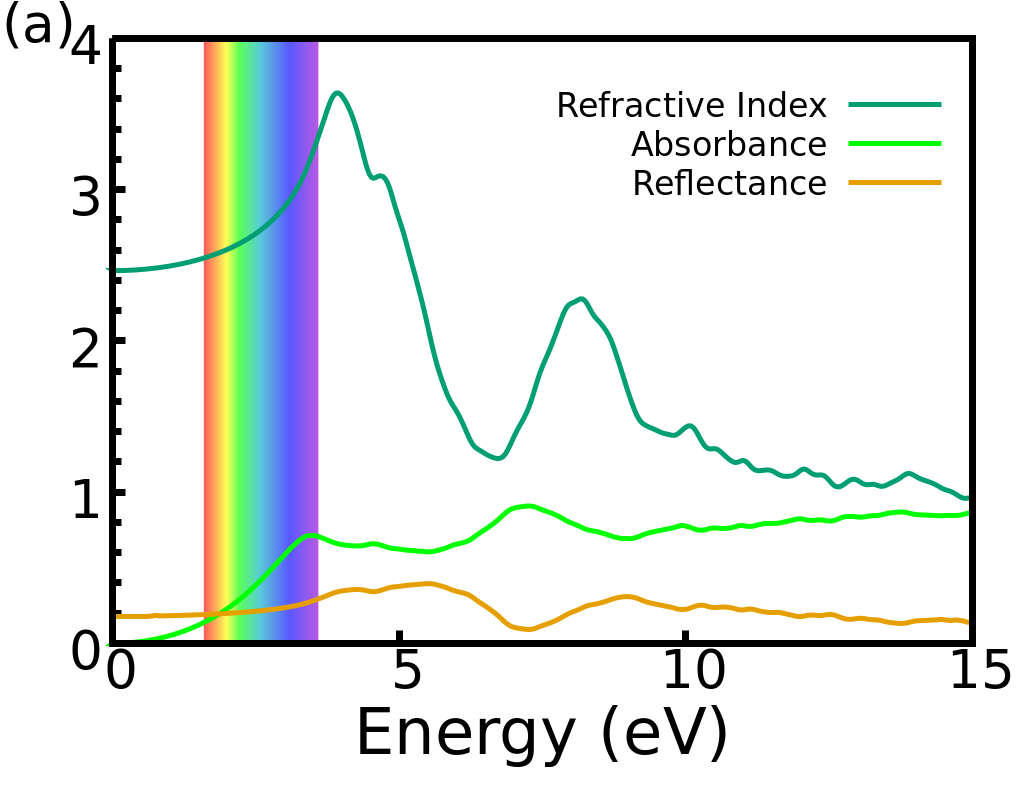}
\includegraphics[height=5cm,width=7cm]{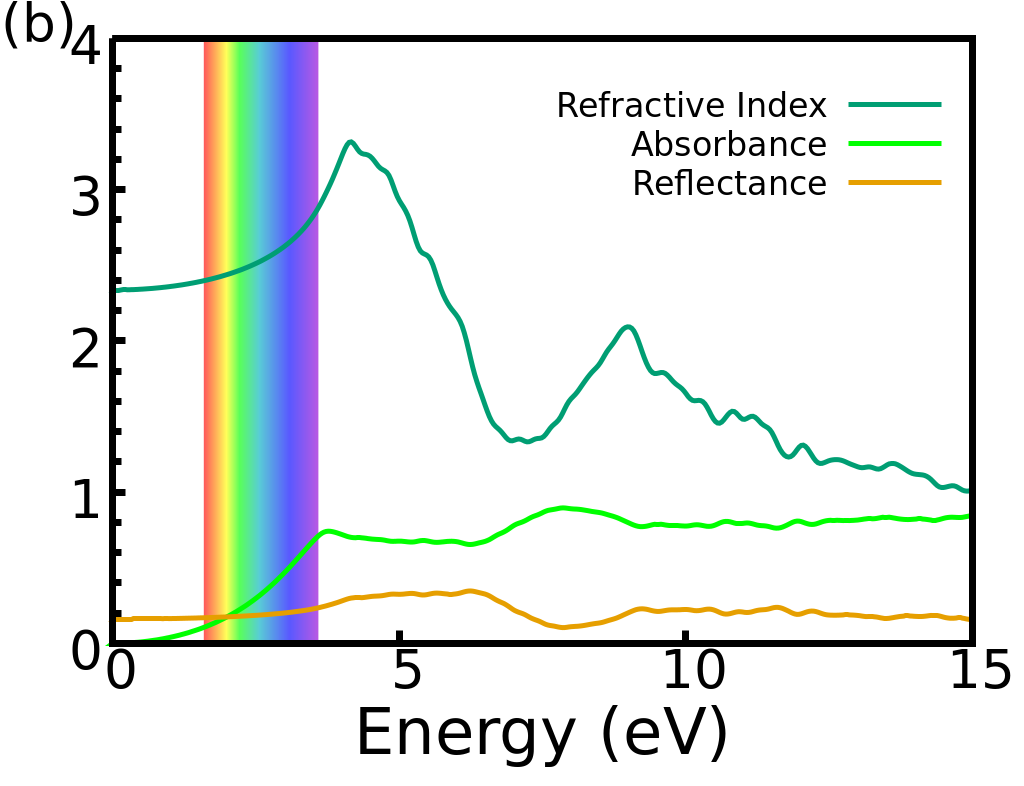}
\includegraphics[height=5cm,width=7cm]{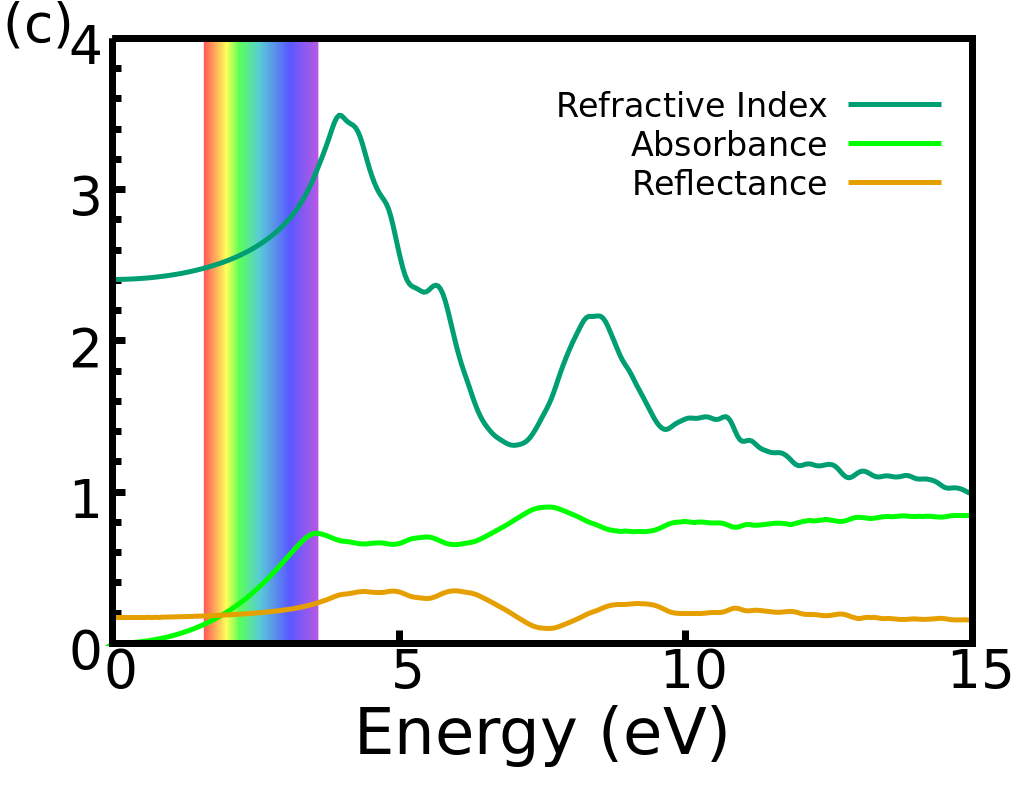}
\caption{Index of refraction, absorbance and reflectance of (a) LNO, (b) LTO and (c) LNTO respectively}
\label{optic_n,A,R}
\end{figure}
\begin{figure}[!h]
\includegraphics[height=5cm,width=7cm]{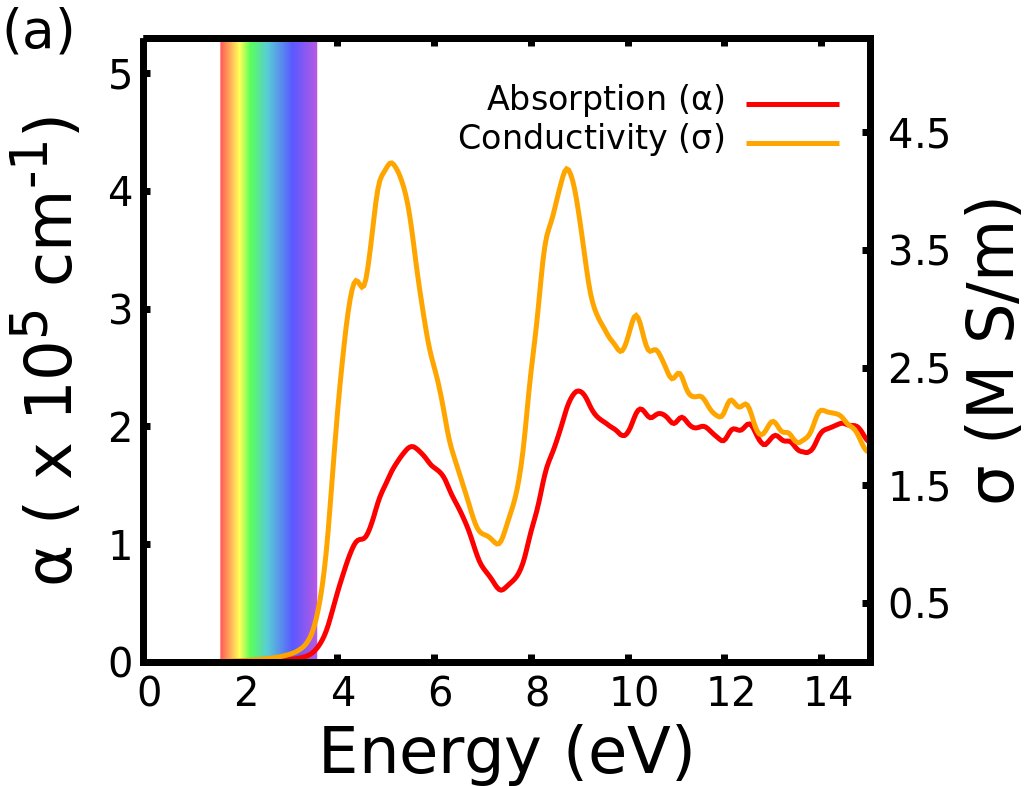}
\includegraphics[height=5cm,width=7cm]{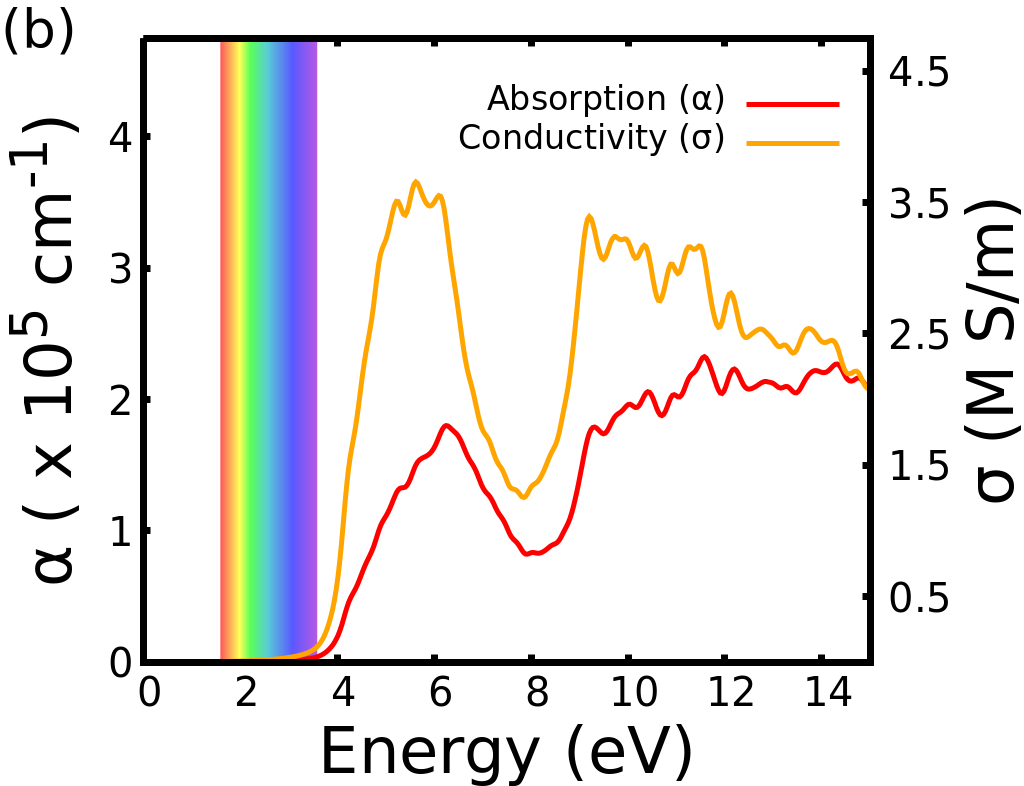}
\includegraphics[height=5cm,width=7cm]{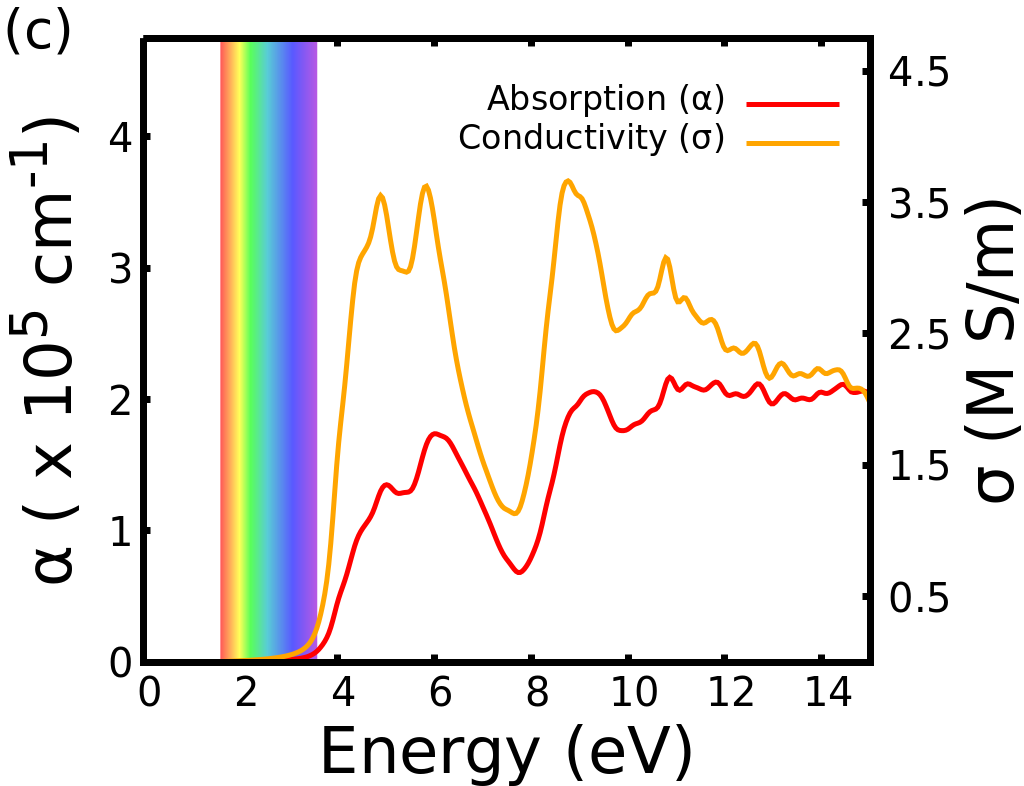}
\caption{Absorption and Conductivity of (a)LNO, (b)LTO and (c)LNTO respectively}
\label{optic_abs_cond}
\end{figure}
The real part shows peaks around 3.85, 3.94 and 4.12 eV close to the visible range within the certain limitation in DFT for LTO, LNTO and LNO respectively and then decreases with increase in photon energy.
In the zero-frequency limit the static dielectric constant $\epsilon_{r}(0)$ values for LNTO is 5.8 which lies in between the value of LNO (6.07) and LTO (5.45) (see Fig. \ref{optic_dielectric}).
The LNTO structure shows a refractive index of 2.40 which lies in between the refractive indices of LNO (2.46) and LTO (2.33) [Fig. \ref{optic_n,A,R}]. The absorption coefficient $(\alpha)$ calculated from the imaginary part of the refractive index which gives the information about how much light is absorbed as it travels through a material.
All the materials show significant absorbance in the visible range. Fig. \ref{optic_n,A,R} indicates around $17\%$ reflectance at 0 eV and it reduces to as less as $10\%$ for all structures as the photon energy increases. The calculated absorption coefficient $\alpha(\omega)$ indicates that it begins to absorb incident light at photon energies of 3.1 eV for LNO and 3.53 eV for LTO, and for LNTO 3.36 eV. The corresponding optical conductivity is also shown in Fig. \ref{optic_abs_cond}.
\subsection{Piezoelectric properties and Born effective charges}
Recent advancements in smart device technology have increasingly leveraged the enhanced piezoelectric properties of various materials, contributing significantly to improved performance and functionality \cite{Akdogan05,Bansevicius11}. 

In a strain-free condition with minimum stress, piezoelectric tensor components are calculated for both clamped and relaxed ion configuration.
$\vert e_{ij}\vert_{max}$ for LNO and LNTO are 3.97 and 3.23 $C/m^{2}$ respectively for LTO it is smaller 2.13 $C/m^{2}$ as the distortion is slightly higher in case of LNO and LNTO than LTO. Both LNO and LNTO possess comparable value with PZT (3.648 $C/m^{2}$) making it suitable for commercial use \cite{debi2025}. 
Off-centre displacement of Nb and Ta leads to deformed octahedra in the perovskite and that increases the piezoelectric property \cite{Prem21}. 
This deformation arises due to two factors, smaller radius of Li and strong covalency of Nb and Ta atoms with oxygen. Since the polarisation caused by this deformation depends on the covalent strength of $B$-site polyhedra, where Nb-O is more covalent than Ta-O and we observe corresponding piezoelectric behaviour. In LNTO, having both elements in equal proportion gains a value in between them. Because of the off-cubic structures, diagonal traits are not visible in the piezoelectric tensor matrices for both clamped and relaxed ion calculations. 

Born effective charges (BECs) quantifies the polarisation behaviour with respect to atomic displacements and can be defined as $Z_{\alpha\beta}^{*}=\Omega\dfrac{\partial P_{\alpha}}{\partial u_{\beta}}$ where, $P_{\alpha} $ represent polarisation in $\alpha$ direction per unit volume, $u_{\beta}$ is the atomic displacement along $\beta$ direction and $\Omega$ is the total volume \cite{Ravindran06}.
 It is observed that Li has BEC close to nominal value showing minimal contribution to polarisation. The BEC value of both Nb and Ta is $> 7.3 e$ in all the compounds and for Ta the value is slightly higher than Nb because of the higher shielding effect. 
  Large diagonal component signifies strong coupling between atomic displacement and polarisation which is essential for ferroelectrics and piezoelectrics \cite{Hong24, Gonze97}. Hybridisation of p-d orbitals leads to strong covalency inducing large dipole moments. This results high piezoceramic value of LNO rather than LTO despite a high BEC of $7.41 e$ for Ta and $7.27 e$ for Nb in LTO and LNO respectively. Maintaining the average property of both perovskite LNTO have BEC $7.33 e$ corresponding to Ta atom. BEC values of oxygen are $-4.06 e$, $-3.98 e$ and $-3.57 e$ for LNO, LTO and LNTO proving the explanation of covalency.

\section{Conclusion}
A comprehensive investigation of the electronic structures, thermal, optical and piezoelectric  properties of lithium niobate, lithium tantalate, and their double perovskite counterpart LNTO is carried out using DFT, DFPT, and TD-DFT. 
The calculated electronic band gap ranges from $3.5$ to $3.7$eV, aligning well with reported results. The density of states reveals that the conduction band is primarily governed by the transition metal $d$-orbital, whereas the valence band is dominated by O-$2p$ state. The phonon dispersion curves confirm the dynamical stability of all three systems, with acoustic modes driven by the heavier atoms in the lattice. Thermal analysis indicates that LTO possesses the lowest free energy, making it the most thermally stable among the three, followed by LNTO and then LNO. The optical properties demonstrate significant activity near the UV edge of the visible spectrum, with LNO exhibiting superior absorption, static dielectric constant and optical conductivity compared to LNTO and LTO. The piezoelectric coefficients of 3.97 $C/m^{2}$ and 3.23 $C/m^{2}$, which are comparable to that of PZT (3.64 $C/m^{2}$), highlight the enhanced piezoelectric response of these materials. The strong covalent bonding between $2p-4d$ or $2p-5d$ orbitals leads to substantial spontaneous polarization, effectively compensating for structural deformation caused by the smaller Li$^{+}$ cation at the A-site. The octahedral distortion remains stable, imparting remarkable optical and piezoceramic characteristics. The Born effective charge (BEC) values reflect the influence of lattice deformation and covalency, with Ta exhibiting a larger BEC than Nb, while the bonded oxygen atoms follow the same trend. 
Overall, LNO demonstrates superior optical and piezoelectric performance, whereas LTO exhibits greater BEC values and thermal stability. LNTO, being intermediate in both electronic and physical characteristics, emerges as a promising candidate for diverse device applications.

%
\section*{Declaration of competing interest}
The authors declare that they have no known competing financial interests or personal relationships that could have appeared to influence the work reported in this paper.
\section*{Data availability}
No data was used for the research described in the article

\section*{Acknowledgements}
The authors acknowledge the High Performance Computing (HPC) facility provided by the Center of Excellence in High Energy and Condensed Matter Physics, Department of Physics, Utkal University, India. DP acknowledges the fellowship from RUSA, Utkal University, India.

\bibliography{Reference}

\end{document}